\documentclass[%
 aps, twocolumn, amsmath,amssymb, floatfix,%
]{revtex4-1}

\usepackage{graphicx}% Include figure files
\usepackage{dcolumn}% Align table columns on decimal point
\usepackage{bm}

\usepackage[utf8]{inputenc}
\usepackage[T1]{fontenc}
\usepackage{mathptmx}

\usepackage{xcolor}
\usepackage{ulem}

\begin{document}

\preprint{AIP/123-QED}

\title[Simple mechanism that breaks the Hall effect linearity at low temperatures]{Simple mechanism that breaks the Hall effect linearity at low temperatures}

%Sample Title:\\with Forced Linebreak

% Force line breaks with \\

\author{A. Yu. Kuntsevich}%
 \email{alexkun@lebedev.ru}
 \affiliation{P.N. Lebedev Physical Institute, Russian Academy of Sciences, Moscow 119991, Russia}%Lines break automatically or can be forced with \\
 \affiliation{National Research University Higher School of Economics, Moscow 101000, Russia}
\author{A. V. Shupletsov}%
\affiliation{P.N. Lebedev Physical Institute, Russian Academy of Sciences, Moscow 119991, Russia}%\\This line break forced with \textbackslash\textbackslash
\author{A. L. Rakhmanov}
\affiliation{Institute for Theoretical and Applied Electrodynamics, Russian Academy of Sciences, Moscow 125412, Russia}%

\date{\today}% It is always \today, today,

\begin{abstract}
Hall resistance $R_{\textrm{xy}}$ is commonly suggested to be linear-in-magnetic-field $B$, provided the field is small. We argue here that at low temperatures this linearity is broken due to weak localization/antilocalization phenomena in inhomogeneous systems, while in a uniform medium the linear-in-field dependence of $R_{\textrm{xy}}(B)$ {is preserved}. We calculate the Hall resistance for different two-component media using a mean-field approach and show that this non-linearity is experimentally observable.
\end{abstract}

\maketitle

%\section{\label{sec:level1}First-level heading:\protect\\ The line break was forced \lowercase{via} \textbackslash\textbackslash}

{Ordinary Hall effect} is broadly used to calculate the charge carriers density $n$ since commonly accepted that {Hall resistance} $R_{\textrm{xy}}\propto 1/n$. The latter relation assumes that $R_{\textrm{xy}}$ depends linearly on the value of the transverse magnetic field $B$ and the carrier density $n$ is {sometimes} estimated at %some 
fixed small magnetic field~\cite{Hurd1972}. In other words, one assumes that the Hall coefficient $R_\textrm{H}=R_{\textrm{xy}}/B$ is independent of $B$ in the low magnetic field range. However, there are a number of observations that the Hall coefficient at low temperatures and at low magnetic fields depends on $B$ even in the systems without magnetic impurities~\cite{Newson1987, Ovadyahu1988, Zhang1992, Minkov2010, Joshua2012, Kuntsevich2013}. This means, in particular, that the value of the charge carriers density determined using Hall resistance may be misleading. Several microscopic models were suggested to explain the $R_\textrm{H}(B)$ dependence in the low magnetic field range. In Ref.~\onlinecite{Minkov2010}, this was attributed to the higher-order corrections to the Drude conductivity in $(k_Fl)^{-1}$ (here $k_F$ is the Fermi momentum and $l$ is a mean free path). The memory effect in the electron scattering could also be a reason for the dependence of $R_\textrm{H}$ on $B$ in low field, as it has been shown in Ref.~\onlinecite{Dmitriev2008}. A non-linearity in $R_{\textrm{xy}}(B)$ due to superconducting fluctuations was proposed in Ref.~\onlinecite{Michaeli2012}. 

In this paper we suggest a different mechanism of $R_{\textrm{xy}}(B)$ non-linearity for two-dimensional (2D) systems, which also can be valid for 3D systems close to the metal-insulator transition. This mechanism is simple and rather general. We argue that due to tensor nature of the magnetoresistivity, the observed non-linearity directly follows from the weak localization/antilocalization (WL/WAL) if the system is inhomogeneous {and the spatial scale of inhomogeneties exceeds all WL/WAL lengths.}

WL and {WAL} phenomena, that is, quantum interference effects, lead to a steep low-field magnetoresistance. Let us consider a 2D {homogeneous} isotropic system in the limit of low temperatures {($l_{\varphi}>l$, where $l_{\varphi}$ is the phase-breaking length)} and low transverse magnetic field ($B \ll \hbar/el^2 < 1/\mu$, where $\mu$ is the mobility measured in the inverse Tesla). According to Hikami-Larkin-Nagaoka formula~\cite{Hikami1980}, magnetoconductivity due the {WL} [that is, $\Delta\sigma(B)=\sigma(B)-\sigma(0)$] for such system can be expressed as
\begin{equation}
\Delta\sigma(B)=\alpha \frac{G_0}{\pi} \left[\psi\left(\frac{1}{2}+\frac{\hbar}{4eBl_\varphi^2}\right)-\ln\left(\frac{\hbar}{4eBl_\varphi^2}\right)\right].
\label{HLN}
\end{equation}
Here {$G_0 = e^2/2\pi\hbar$ is the conductivity quantum,} $\psi$ is the digamma function, {$e>0$} denotes elementary charge, and $\alpha$ is a constant typically from $-1$ to 1. {Most commonly in 2D systems electron-electron dephasing mechanism is dominant that leads to $l_\varphi \propto \sqrt{1/T}$~\cite{Narozhny2002}. In 3D systems and thin films the dominant dephasing mechanism is electron-phonon interaction with} {$l_\varphi\propto T^{-\nu}$, where an exponent $\nu$ depends on disorder~\cite{lin}.} 

In Refs.~\onlinecite{Fukuyama1980, Altshuler1980} it was shown that the quantum correction Eq.~\eqref{HLN} does not contribute to the Hall effect, i.e. off-diagonal terms of the resistivity tensor are independent of $\Delta\sigma(B)$. In particular, in the 2D case the corresponding resistivity tensor has a form
\begin{equation}
    \hat\rho=   \left(
                \begin{array}{cc}
                    1/[\sigma(0)+\Delta\sigma]  &   -B/ne   \\
                    B/ne    &   1/[(\sigma(0)+\Delta\sigma] \\
                \end{array}
                \right),
\label{mrtensor}
\end{equation}
where and $\sigma(0)=ne\mu$. {In these terms, the mobility $\mu = \pm e\tau/m^*$ and the charge density $n$ are sign-dependent (negative for electrons), whereas mean free time $\tau$ and effective mass $m^*$ are always positive.} We may introduce a modified field-dependent mobility

\begin{equation}
    \tilde{\mu}(B)=\mu+\Delta\sigma(B)/ne,
    \label{mutilde}
\end{equation}
and write down the conductivity tensor (inverted resistivity) in a standard Drude form

\begin{equation}
    \hat\sigma = \frac{ne\tilde{\mu}} {1 + \tilde{\mu}^2 B^2} \left(
        \begin{array}{cc}
            1               & \tilde{\mu} B \\
            -\tilde{\mu} B  & 1             \\
        \end{array}
    \right).
\label{mctensor}
\end{equation}

We consider here a 2D inhomogeneous medium. For simplicity, we assume that it consists of two species (or phases) of a 2D electron gas with different densities and mobilities, $(n_1,\tilde{\mu}_1)$ and $(n_2,\tilde{\mu}_2)$. The mobilities $\tilde{\mu}_1$ and $\tilde{\mu}_2$ {typically} vary differently with the magnetic field. Therefore, the transport current redistributes between these species when the applied magnetic field changes. As a result, the Hall coefficient becomes field-dependent. {General} solution of the problem of a current flow redistribution even in a two-component medium is rather complicated. However, in some special cases an exact or approximate analytical result can be obtained. 

\begin{figure}
    \includegraphics[width=250pt]{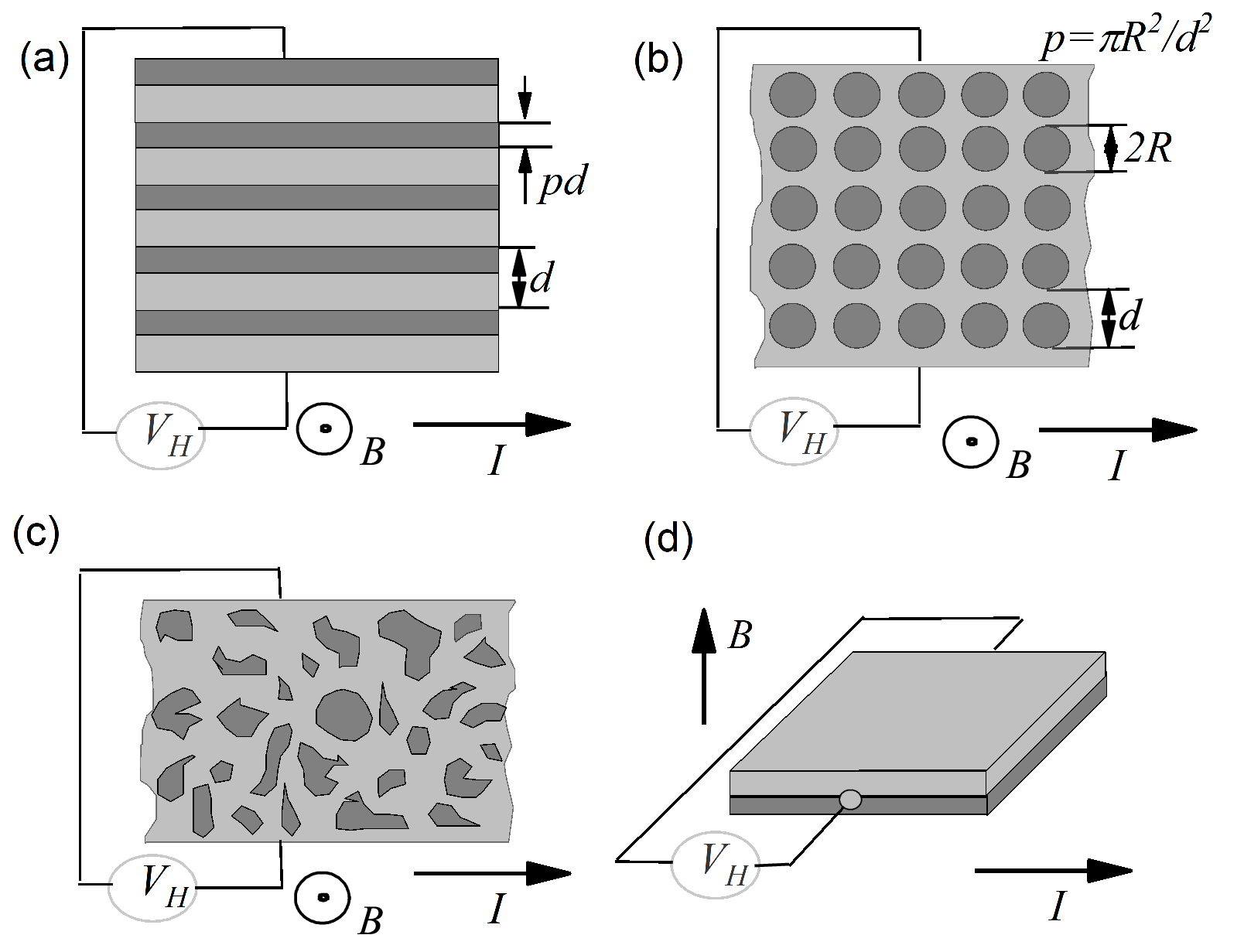}
    \caption{Various geometries of {inhomogeneous} systems, discussed in the paper. The arrows indicate direction of the transport current flow ($I$), voltmeter measures Hall voltage $V_H$. Components with different properties are indicated by gray and light gray colors. {The} fraction of {the dark} gray part in panels (a-c) is equal to $p$.}
    \label{fig:geometries}
\end{figure}

The simplest case of the inhomogeneous system is an array of strips {with} the current flow{ing} parallel to them, as shown in Fig.~\ref{fig:geometries}a. The electric field along the current flow is the same in both components of the system (dark and light gray strips). Summing up Hall voltages for all strips and dividing the result by the total current, we get

\begin{equation}
    \label{Strips}
    R_{\textrm{xy}} = \frac{B}{e} \cdot \frac{p\tilde{\mu}_1 + (1-p)\tilde{\mu}_2} {pn_1\tilde{\mu}_1 + (1-p)n_2\tilde{\mu}_2}.
\end{equation}
Here $p$ denotes the fraction of the dark gray component ($n_1,\tilde{\mu}_1$). According to this formula, the Hall coefficient $R_\textrm{H}(B)$ is not a constant in the low magnetic field if $\tilde{\mu}_1$ and $\tilde{\mu}_2$ {are not proportional to each other}.

Another model system that we consider here is a regular array of circular inclusions in the conductive matrix, as shown in Fig. \ref{fig:geometries}b. {In the inclusions} the carrier density is $n_1$ and mobility is $\tilde{\mu}_1$, while in the matrix these values are $n_2$ and $\tilde{\mu}_2$, {respectively}. To obtain an approximate solution to this model, we apply a self-consistent mean-field theory~\cite{LL,MFT}. In this approach each element of the array {is approximated} as a circular unit consist{ing} of the inclusion in the center { and} the matrix shell. We place this unit in the media with an effective conductivity tensor, solve corresponding electromagnetic problem, average the calculated electric field self-consistently, and obtain mean-field equations for the effective conductivity. Details of such calculations are described in Appendix~\ref{app1}. The obtained result can be presented in the form

\begin{equation}
    \label{islands}
    R_{\textrm{xy}} = \frac{B} {n_2e} \cdot \frac{S^2 + 2p [2n_1 n_2 \tilde{\mu}_1 (\tilde{\mu}_1 - \tilde{\mu}_2) - D^2] + p^2 D^2} {(S - p D)^2},
\end{equation}
where $D = n_2 \tilde{\mu}_2 - n_1 \tilde{\mu}_1$, and $S = n_1 \tilde{\mu}_1 + n_2 \tilde{\mu}_2$. %The mobilities in the circular islands, $\tilde{\mu}_1$, and in the remaining 2D gas, $\tilde{\mu}_2$, are calculated using Eq.~(\ref{mutilde}).

The same mean-field approach {can be applied} to calculate the effective conductivity tensor in the case of random mixture of two 2D electron phases. We approximate the regions with different conductivities by a circular inclusions with different radii. The analytical solution of the problem is tremendous and we present it only in Appendix~\ref{append2} with the details of calculations. {In a random mixture the inclusions join in clusters, the characteristic size of these clusters increases with the increase of $p$ and becomes infinite at some $p=p_c$, which is called percolation threshold.} The mean-field theory usually gives a good result, when the fraction content $p$ is not close to $p_c$~\cite{LL,MFT}. For isotropic 2D systems $p_c=0.5$, while for isotropic 3D structures $p_c\approx 0.15$. 

Now let us analyze a possible {amplitude} of the Hall resistance non-linearity due to system inhomogeneity. This non-linearity arises due to the {either} difference in the carriers' mobility in different parts of the sample {or due to difference in the corresponding timescales (phase breaking time and spin-orbit interaction time)}. The larger is this difference, the greater would be the non-linearity. Naively, according to Eqs.~(\ref{HLN}) and (\ref{mutilde}) there is no limitation on relative mobility variations with the magnetic field. Indeed, the effect would be high, if the system contains low-$n$ regions, since correction to the mobility is inversely proportional to $n$, see Eq.~(\ref{mutilde}). However, in the low-$n$ regions the conductivity itself is low, and the higher order corrections in $(ne\mu)^{-1}$ come into play, suppressing the magnetoresistance~\cite{Minkov2004}. As a result, a realistic estimate of the relative variation of $\tilde{\mu}$ due to {WL} could be maximum 50\% or so for 2D systems and for 3D systems, as well~\cite{Zhang1992,Belykh2018}.

\begin{figure}[h]
    \includegraphics[width=250pt]{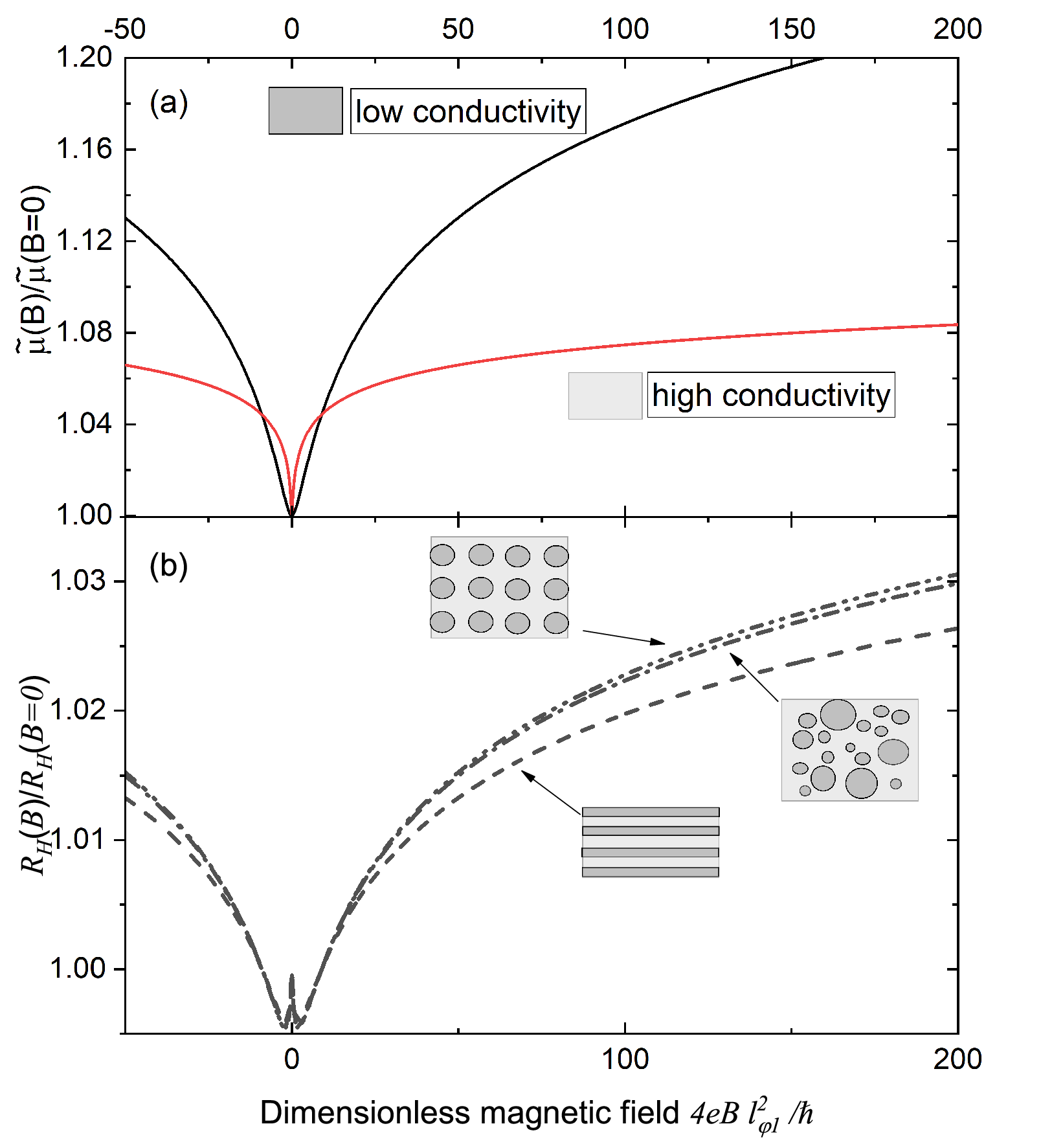}
    \caption{(a) Magnetic field {dependence of the} mobility {$\tilde{\mu}(B)/\mu = 1 + \Delta\sigma(B)/\sigma(0)$} calculated using Eqs.~(\ref{HLN}) and \eqref{mutilde}. (b) Magnetic field dependencies of the Hall coefficient $R_\textrm{H}(B)$ for inhomogeneous systems, calculated for three model systems using the data from panel (a) as an input. The fraction of the low-conductivity phase with $n_1=n_2/5$ is $p=0.3$. The functions $\tilde{\mu}(B)$ and $R_\textrm{H}(B)$ are normalized to its values at $B=0$, the field scale is normalized using the phase breaking length $l_{\varphi 1}$ for the low-conductivity phase.}
    \label{fig:models} 
\end{figure}

Note here, that Eq. (\ref{HLN}) describes magnetic-field induced dephasing in the diffusive limit of long interference loops. This formula is valid only in the low magnetic field limit $B\ll \hbar/(el^2) < 1/\mu$. In higher magnetic fields, $B\sim \hbar/(el^2)$, the logarithmic asymptotic of Eq. (\ref{HLN}) should be replaced by $\propto 1/\sqrt{B}$~\cite{Kawabata1984, Dmitriev1997}. In this regime, called ballistic, there is no a simple analytical expression for the magnetoconductivity. Below, we use Eqs.~(\ref{HLN}) and (\ref{mutilde}) for the qualitative analysis of the non-linear Hall effect.

The magnetic field dependence of the mobility and the Hall coefficient for the outlined above three inhomogeneous systems (parallel {strips}, ordered and disordered circular inclusions)  are illustrated in Fig.~\ref{fig:models}. We assume that at zero magnetic field the mobilities in both fractions are equal ($\mu_1=\mu_2$), while the carrier density are different ($n_1=n_2/5$). This difference in the charge carrier densities gives rise to difference in the phase breaking length $l_\varphi$ {approximately} by a factor of~$5$~\cite{Narozhny2002}. We also put $p=0.3$ for all calculated curves in Fig.~\ref{fig:models}b. 

The effect of the {WL} on the charge mobility in two different fractions of the considered systems is illustrated in Fig.~\ref{fig:models}a. The {whole} variation of the mobility with the magnetic field {is equal to $e^2/2\pi\hbar\sigma$, i.e., higher for }the low-$n$ phase. {At the same time, low field asymptotic for Eq.~\eqref{HLN} is $\Delta\sigma(B)\approx(\alpha/24\pi) G_0 B^2/(\hbar/4el_\varphi^2)^2 \propto B^2(n\mu)^4$. It means that the low-field drop is sharper for high-$n$ phase. These two features result in the intersection of normalized $\tilde{\mu}(B)$ (Fig.~\ref{fig:models}a) 
leading to nonmonotonic $R_H(B)$ behavior}, as illustrated in Fig.~\ref{fig:models}b. 
%of two phases at some field $B_{cross}$. Thus, the current density increases with the increase of the magnetic field in the high-$n$ phase till $B\sim B_{cross}$ and decreases in the low-$n$ phase.} The current flow pattern varies giving rise to the dependence of the Hall coefficient on the magnetic field, as illustrated in Fig.~\ref{fig:models}b. 

As it is seen from Fig.~\ref{fig:models}b, %{the magnetic field dependence of the  Hall coefficient has a sharp peak near zero field and increases smoothly at higher $B$. These non-monotonic behavior directly follows from the crossover between $\tilde{\mu}(B)$ in the two phases.} 
{the high-field $R_H(B)$} increase is higher for the systems with the circular inclusions than for the parallel strips. This behavior has a simple physical explanation. In the system with parallel strips the transport current mainly passes through the highly-conductive strips since they occupy 70\% of the sample and their conductivity is five times higher. As a result, the increase of the conductivity of low-conductive phase with the growth of $B$ results in a slight redistribution of transport flow. In the system with circular inclusions, the transport current flows mainly around the low-$n$ inclusions. This way is long, and as the conductivity of the low-$n$ phase regions drops with magnetic field, the path though them becomes more preferable, the fraction of transport current in the low-$n$ islands increases giving rise to a more pronounced change in the Hall coefficient. The effect is the strongest in the structure with ordered circular inclusions.  

A similarity of the obtained results for three different model systems implies that the particular type of inhomogeneity is not crucial for a qualitative picture of the effect.

In the special case $p=0.5$, the Hall resistance in Eq. (\ref{Strips}) is similar to that for a two-liquid model. Within this model, the {system is supposed to consist of} two different types of charge carriers with different conductivities $\hat\sigma^{(1)}$ and $\hat\sigma^{(2)}$. The conductivity tensor of the system is a sum $\hat\sigma=\hat\sigma^{(1)}+\hat\sigma^{(2)}$. The two-band model is used to describe multiband systems, e.g., doped topological insulators~\cite{taskin2011}, semimetals~\cite{kvon2008, zhou2016}, etc. A similar model can be also applicable for a bilayer shown schematically in Fig.~\ref{fig:geometries}d. A question is, why the non-linearity of the Hall resistance is not observed in all inhomogeneous and multicomponent electron systems? We believe that a crucial factor is a scattering between different types of charge carriers. Indeed, as it was shown for multiband compounds~\cite{Averkiev2001}, multivalley systems~\cite{Kuntsevich2007, McCann2006}, and topological insulators~\cite{Brahlek2014}, if we take into account the scattering between different types of the quasi-particles, a multicomponent system effectively reduces to a single-component one from the {WL/WAL} point of view.

The macroscopic spatial separation of different phases guarantees that the {WL} occurs in them independently justifying our approach. If a typical scale of the inhomogeneities is comparable or smaller than the phase breaking length $l_\varphi$, the {WL} in different parts of the inhomogeneous system can not be considered as independent, since the charge carrier passes through the regions with different mobilities and electron densities during dephasing time. This effect results in the diminishing of the corrections to the Hall coefficient due to the system inhomogeneity {and becomes} especially important when $T\rightarrow 0$. However, further microscopic study of the {WL} is necessary since dephasing itself depends on the system inhomogeneity~\cite{Germanenko2001}.  

\begin{table*}[ht]
    {    \caption{\small Magnetic field and transport scales of WL from representative experiments\cite{WLGraphene_1,WLGraphene_2,WLGaAs,WLInGaAs} and theoretical estimates (last column): $B_{\varphi} = \hbar/4el_{\varphi}^2$ is a crossover between low-field parabolic and high-field logarithmic dependencies; $l_{\varphi}$ is taken as $l\sqrt{\tau_{\varphi}/2\tau}$, where $\tau^{-1}_{\varphi} = (k_B T/\hbar) (2G_0/\sigma) \ln(\sigma/2G_0
    )$~\cite{Narozhny2002};
    transport field $B_{tr} = \hbar/2el^2$ denotes the transition between the diffusive ($B<B_{tr}$; logarithmic behaviour) and ballistic ($B>B_{tr}$; $1/\sqrt{B}$ asymptotic) regimes; $1/\mu$ is the transition between classically weak and strong magnetic fields. Effective mass $m^*$ is $0.067m_0$ for GaAs, and $0.058m_0$ for In$_{0.2}$Ga$_{0.8}$As ($m_0$ is free electron mass); for graphene  $m^* = p_F/v_F = \hbar\sqrt{2\pi n/g}/v_F$ where $v_F = 10^6$~m/s. $g$ denotes valley degeneracy ($g=1$ for GaAs and InGaAs and $g=2$ for graphene).}
    \begin{tabular}[t]{|c|c|c|c|c|c|}
        \hline
        Parameter   & Graphene \cite{WLGraphene_1}  & Graphene \cite{WLGraphene_2}      & GaAs \cite{WLGaAs}    & In$_{0.2}$Ga$_{0.8}$As \cite{WLInGaAs}    & Theoretical\\
                       & $n=3\cdot10^{12}$ cm$^{-2}$   & $n=2.5\cdot10^{11}$ cm$^{-2}$     & $n=7.15\cdot10^{11}$cm$^{-2}$ & $n=1.1\cdot10^{12}$cm$^{-2}$   & estimate  \\
                       & $\mu=0.395$~m$^2$/Vs          & $\mu=1.34$~m$^2$/Vs               & $\mu=0.49$~m$^2$/Vs         & $\mu=0.094$~m$^2$/Vs           &           \\
        %\hline
        %$T_{\rm onset}$ [K]     &   $\displaystyle\sim\dfrac{h^2}{8\pi k_B} \dfrac{n}{m^*\ln{\dfrac{hn\mu}{2e}}}$
        %                        &   $\approx 625$   
        %                        &   $\approx 260$
        %                        &   $\approx 90$
        %                        &   $\approx 350$               \\
        \hline
        $\sigma$ [$e^2/2\pi\hbar$]      
                                &   49.1
                                &   13.9
                                &   14.6
                                &   4.3
                                &   $\dfrac{2\pi\hbar}{e}n\mu$  \\
        \hline
        $B_\varphi$[T] at $T=$2K
                                &   $\approx 8\cdot10^{-5}$
                                &   $\approx 2.1\cdot10^{-4}$
                                &   $\approx 8.1\cdot10^{-4}$
                                &   $\approx 3.4\cdot10^{-3}$
                                &   $\displaystyle\sim\frac{ge}{4\pi^2\hbar^3} \frac{m^*}{n^2\mu^2}\ln\left(\frac{\pi\hbar n\mu}{e}\right)k_BT$ \\
        \hline
        $l_\varphi$[$\mu$m] at $T=$2K
                                &   $\approx1.41$
                                &   $\approx0.89$
                                &   $\approx0.45$
                                &   $\approx0.22$
                                &   $\displaystyle\sim\frac{\pi\hbar^2}{e} n\mu \sqrt{\frac{1}{gm^*\ln{\frac{\pi\hbar n\mu}{e}}}} \frac{1}{\sqrt{k_BT}}$    \\
        \hline
        $B_{tr}$[T]             
                                &   0.05
                                &   0.05
                                &   0.07
                                &   1.3
                                &   $\dfrac{\hbar}{2el^2}=\dfrac{ge}{4\pi\hbar}\dfrac{1}{n\mu^2}$   \\
        \hline
        $l$[nm]                 
                                &   80
                                &   78
                                &   69
                                &   16
                                &   $\displaystyle\frac{\hbar}{e}\mu\sqrt{\frac{2\pi n}{g}}$    \\
        \hline
        $1/\mu$[T]              
                                &   2.5
                                &   0.7
                                &   2
                                &   10.6                       
                                &   $1/\mu$ \\
        \hline
        $l_\varphi/l$ at $T=$2K 
                                &   $\approx 17.7$
                                &   $\approx 11.4$
                                &   $\approx 6.6$
                                &   $\approx 13.4$               
                                &   $\displaystyle\sim\hbar \sqrt{\dfrac{\pi n}{2m^*\ln{\frac{\pi\hbar n\mu}{e}}}} \frac{1}{\sqrt{k_BT}}$   \\
        \hline
    \end{tabular}
    \label{tablecompar}
}
\end{table*}

Another reason why there are no many reported manifestations of the magnetic field dependence of the Hall coefficient is a so-called ``a textbook paradigm'', which unequivocally affirms that the low-field Hall voltage is linear in $B$~\cite{Hurd1972}. Common methods of the Hall effect study in low magnetic field include: measurements in a fixed field $\pm B$ with subsequent antisymmetrization of the results; a sample  rotation in a constant magnetic field~\cite{Hermann1965}; low-amplitude AC-technique with subsequent averaging the signal over a small-field range~\cite{Lupu1967}; a simple linear extrapolation of $R_{\textrm{xy}}(B)$ dependence. It is worth to mention, that the magnetic field dependence of $R_{\textrm{xy}}(B)$ is visually indistinguishable from the straight line even if $\tilde{\mu}$ changes by several tens \%. 
As a result, the low-field Hall effect non-linearity was reported in a few experiments when this effect was looked for intentionally\cite{Newson1987, Ovadyahu1988, Zhang1992, Minkov2010, Joshua2012, Kuntsevich2013}. One of the goals of this paper is to motivate experimentalists to look for the low magnetic field non-linearity in the Hall effect.

{In Table~\ref{tablecompar} we present magnetic field scales and transport characteristics  experimentally observed for 2D electron gas in graphene~\cite{WLGraphene_1,WLGraphene_2},  AlGaAs/GaAs/AlGaAs~\cite{WLGaAs}, and GaAs/InGaAs/GaAs~\cite{WLInGaAs} quantum wells. Given numbers indicate that the values of WL-related parameters may vary in a wide range. Also Table~\ref{tablecompar} contains the theoretical formulas for the experimentally observed quantities. As seen from these estimates, these observables may by tuned strongly by disorder, charge carriers density, and temperature.
}

{Note that }the above considerations are not applicable to the Nernst effect - a ``thermoelectric brother'' of the Hall effect - since the correction due to {WL} to the Nernst coefficient is significant even in the homogeneous system~\cite{NernstEffect}.

The discussed above mechanism of the Hall effect non-linearity is simple and robust against the system specific details. We believe that it should be widely observed. The main idea is equally applicable to {WL} and {WAL} in 2D and 3D systems. In general, the carrier density fluctuations exist in any system. In order to observe Hall resistance non-linearity due to intrinsic disorder, two conditions should be fulfilled: (i) the spatial scale of the inhomogeneity should be larger than the phase breaking length, (ii) phase breaking length should be larger than electron mean free path. 

{It is worth to mention that the Hall-effect nonlinearity in the inhomogeneous systems is not {limited by WL/WAL} and may arise due to any magnetoresistance mechanism that nevertheless preserves linear $R_{xy}(B)$ dependence in the homogeneous system.}

The discussed effects could be observed in the systems with a tendency to formation of spatially inhomogeneous state \cite{martin2008,morgun2016,stolyarov2020}, or structurally non-uniform systems, e.g., like mixture of single-layer and bilayer graphene, which is naturally obtained in the chemical vapour deposition growth process~\cite{Sutter2008}.
A promising idea is to prepare a tunable 2D inhomogeneous system using independent gate electrodes controlling different parts of the 2D electron gas~\cite{Kuntsevich2016,Nunuparov}. Tunability of the system components may strongly enhance the effects under discussion. In particular, the {one} can prepare the components with the carriers having different sign of the charge (electron-hole mixture). 
In so doing, according Eqs.~\eqref{mutilde} and (\ref{mctensor}), the {one} can turn to {sign-alternating Hall effect, and in particular to the case}, in which a global current redistribution occurs~\cite{Alekseev2015}. 

We believe that the proposed mechanism might be relevant to explain some  observations of the low-field Hall effect non-linearity. Indeed, in Ref.~\cite{Ovadyahu1988}, this non-linearity was observed in disordered amorphous films of indium oxide, where domains with different properties may form. Refs.~\cite{Newson1987,Zhang1992} are devoted to diluted semiconductors close to metal-insulator transition, where manifestation of the {WL} is especially strong and fluctuations of the dopant concentrations are possible. The sign of the effect in Refs.~\cite{Newson1987,Ovadyahu1988,Zhang1992} agrees with what we expect from Fig.~\ref{fig:models}. At LaAlO$_3$/SrTiO$_3$ interface (see Fig.~2b of Ref.~\cite{Joshua2012}) low-field Hall non-linearity is also observed. The sign of this non-linearity is opposite to that plotted in Fig.~\ref{fig:models}. However, our explanation may remain reasonable since in this material {WAL} was observed instead of {WL} (see Fig.~2c of Ref.~\cite{Joshua2012}). 

{\bf Conclusions.} In this paper we argue that in inhomogeneous systems the weak localization or antilocalization lead to low-field non-linear magnetic field correction to Hall resistance. This effect arises due to transport current redistribution between components of the inhomogeneous system with the change of the magnetic field. The effect is measurable and can be observed in two- and three-dimensional systems.

\begin{acknowledgments}
The authors are thankful to V.~Yu.~Kachorovskii and L.~E.~Golub for discussions. The work is supported by RFBR Grant 18-32-20202.

\end{acknowledgments}

\appendix
\section{Mean field calculations for a regular two-component media}\label{app1}

\begin{figure}[t]
\centerline{\includegraphics[width=0.85\linewidth]{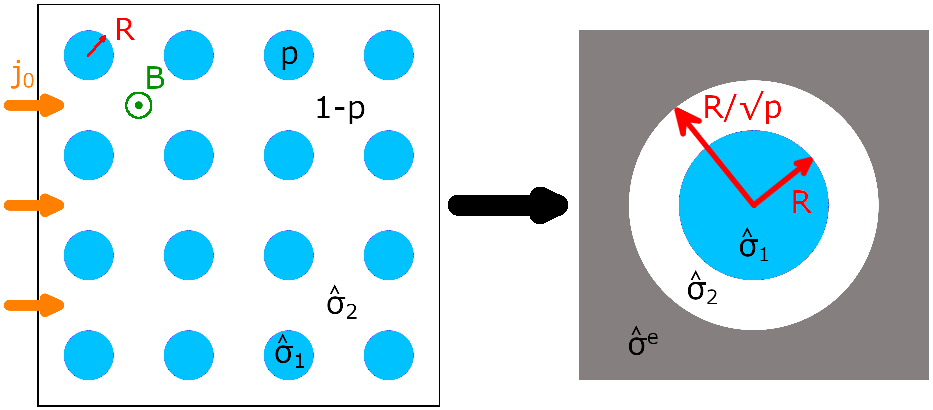}}
\caption{Schematic image of the transition of theoretical model according to mean field theory}
\label{Image}
\end{figure}

An approximate analytical expression for the conductivity of the regular array of equivalent circular islands embedded into a conductive matrix can be obtained analytically by means of effective media approach~\cite{MFT}. We consider infinite 2D array of conductive circular islands with radius $R$ and period $d$, see Fig.~\ref{Image}. The input parameters are magnetic field directed perpendicular to the plane of the sample and the conductivity tensors of the islands $\hat{\sigma}_1$ and the residual 2D electron gas with $\hat{\sigma}_2$. The conductivity tensors have a structure given by Eq.~\ref{mctensor}. We set a boundary condition that DC transport current with given average density $j_0$ flows through the sample. Our goal is to calculate an effective conductivity tensor of the inhomogeneous system $\hat{\sigma}^{e}$.
From the symmetry consideration this tensor must have the following form
\begin{equation}\label{tensorA}
\hat{\sigma}^e=\left(
\begin{matrix}
	\sigma^e_{xx} & \sigma^e_{xy} \\
	-\sigma^e_{xy} & \sigma^e_{xx}
\end{matrix} \right)
\end{equation}

We have two independent quantities $\sigma^e_{xx}$ and $\sigma^e_{xy}$. These values have to be expressed through $\sigma^{(n)}_{xx}$, $\sigma^{(n)}_{xy}$ of islands (n=1) and the remaining 2D gas (n=2), and geometrical factor $p={\pi R^2}/{d^2}$ that denotes the fraction of the system that occupied by the islands.

We treat the periodical system within Wigner-Sietz approach, that is, we replace the square unit sell by a circular one, see left panel in Fig.\ref{Image}. This circular unit cell consists of the island with radius $R$ and conductivity $\hat{\sigma}_1$ in the center and the 2D electron gas shell with radius $R_1=R/\sqrt{p}$ and conductivity $\hat{\sigma}_2$. Following the mean-field approach~\cite{LL,MFT}, we put the system unit cell to the effective media with conductivity tensor $\hat{\sigma}^e$. Then, we solve corresponding electromagnetic problem setting the condition that the current density $j$ far from the center of the circular cell is equal to $j_0$. Finally, we use the self-consistency condition that the average transport current density in the unit cell is equal to $j_0$.         

In each part of our inhomogeneous system the current conservation condition ${\rm div~}\textbf{j}=0$ and Ohm's law $\textbf{j}=\hat{\sigma}\textbf{E}$ are fulfilled. We introduce the electrical potential $\phi$, where $\textbf{E}=-\nabla\phi$. The potential $\phi$ evidently obeys the Laplace's equation 
\begin{equation}\label{Laplace}
\Delta\phi=0,
\end{equation}
where operator $\Delta$ is taken in 2D with the coordinates $x$ and $y$. The solution $\phi$ in different media are matched on the borders using the conditions of continuity of the electrical potential and radial component of the current $j_r$. The solutions must satisfy the boundary conditions $j_x=j_0$ and $j_y=0$ at $x,y\rightarrow\infty$ or in terms of the electrical potential 
\begin{equation}\label{currentA}
\sigma^e_{xx}\frac{\partial\phi}{\partial x}+\sigma^e_{xy}\frac{\partial\phi}{\partial y}\!=\!-j_0,\,\,\,
\sigma^e_{xx}\frac{\partial\phi}{\partial y}-\sigma^e_{xy}\frac{\partial\phi}{\partial x}\!=\!0,\,\,\,
x,y\!\rightarrow\!\infty.
\end{equation}
In the polar coordinates $(r,\theta)$ the solution to Eqs.~\eqref{Laplace} and \eqref{currentA} reads
\begin{eqnarray}\label{solutionA}
\phi&=&r(a\cos{\theta}+b\sin{\theta}),\qquad 0<r<R,\\
\nonumber
\phi&=&\left(d_1r+\frac{d_2}{r}\right)\cos{\theta}+\left(c_1r+\frac{c_2}{r}\right)\sin{\theta},\quad R<r<R_1,\\
\nonumber
\phi&=&\frac{f\cos{\theta}+g\sin{\theta}}{r}-\frac{j_0r(\sigma^e_{xx}\cos{\theta}+\sigma^e_{xy}\sin{\theta})}{\sigma^{e2}_{xx}+\sigma^{e2}_{xy}},\,\,\, r>R_1,
\end{eqnarray}
where eight constants $a$, $b$, $d_i$, $c_i$, $f$, and $g$ are determined from eight continuity conditions of $\phi$ and $j_r=j_x\cos{\theta}+j_y\sin{\theta}$ at the boundaries $r=R$ and $r=R_1$. To express $\hat{\sigma}^e$ through $\hat{\sigma}^{1,2}$ and $p$, we should add self-consistency conditions
\begin{equation}\label{self_consA}
\hat{\sigma}^e\cdot\overline{\textbf{E}}=\left( 
\begin{matrix}
j_0 \\ 0
\end{matrix} \right)
\end{equation}
where $\overline{\textbf{E}}$ is averaged over unit cell electric field
\begin{equation}
\pi R_1^2 \overline{\textbf{E}} = \int^R_0 rdr\int_0^{2\pi}d\theta\textbf{E}(\textbf{r}) + \int^{R/\sqrt{p}}_R rdr\int_0^{2\pi}d\theta\textbf{E}(\textbf{r})
\end{equation}
After rather cumbersome but straight-forward algebra we derive  
\begin{equation}
\label{Expres}
\begin{aligned}
\frac{\sigma^e_{xx}}{\sigma^{(2)}_{xx}}= 	& \left\{ \frac{(1-p)(1+\alpha^2)+\beta p(1-\alpha\gamma)}{(1-p)^2(1+\alpha^2)+\beta p[2(1-p)+\beta p]}+ \right. \\
																			& \left. -\frac{\gamma\sigma^{(2)}_{xy}}{2\sigma^{(2)}_{xx}}-\frac{1}{2} \right\} \frac{2}{1+\gamma^2} , \\
\sigma^e_{xy}=												& \;-\gamma\sigma^e_{xx}, \\
\gamma = 															& -\left\{ \frac{\sigma^{(2)}_{xy}}{2\sigma^{(2)}_{xx}}+\frac{\beta p\alpha}{(1-p)^2(1+\alpha^2)+\beta p[2(1-p)+\beta p]} \right\}\cdot \\
																			& \cdot \left\{ \frac{(1-p)(1+\alpha^2)+\beta p}{(1-p)^2(1+\alpha^2)+\beta p[2(1-p)+\beta p]}-\frac{1}{2} \right\}^{-1}
\end{aligned}
\end{equation}
where
\begin{equation}
\label{Express2}
\begin{aligned}
\sigma^{(i)}_{xx}=\frac{n_i\tilde{\mu}_ie}{1+(\tilde{\mu}_i B)^2}, \qquad &
\sigma^{(i)}_{xy}=\frac{n_i{\tilde{\mu}_i}^2 Be}{1+(\tilde{\mu}_i B)^2}, \\
\alpha=\frac{\sigma^{(1)}_{xy}-\sigma^{(2)}_{xy}}{\sigma^{(1)}_{xx}+\sigma^{(2)}_{xx}}, \qquad &
\beta=\frac{2\sigma^{(2)}_{xx}}{\sigma^{(1)}_{xx}+\sigma^{(2)}_{xx}}
\end{aligned}
\end{equation}

In the low-field limit, $\tilde{\mu}_1B, \tilde{\mu}_2B \ll 1$, we can neglect quadratic in $B$ terms, and obtain the Hall resistance in the low temperature and low field limit $R_{\textrm{xy}}=-\gamma/{\sigma_{xx}}^e$, where $R_{\textrm{xy}}$ obeys Eq.~(\ref{islands}).

\section{Mean field calculations for a random two-component media}\label{append2} 

We consider an isotropic random two-component mixture of phases with different conductivity tensors $\hat{\sigma}^{(1)}$ and $\hat{\sigma}^{(2)}$. We approximate the inclusions of different phases by rings with different radii $R$. Following the mean-field approach~\cite{LL,MFT}, we consider the inclusion with a radius $R$ with $\hat{\sigma}^{(i)}$ (where $i=1,2$) placed in the matrix with the effective conductivity $\hat{\sigma}^e$. We solve corresponding electric problem and, then, we average the electric field over the sample volume and obtain the self-consistency conditions, similar to that performed in Appendix~\ref{app1}. Note, the present calculations can be easily generalized on the case of several components with different conductivities. 

The solution for the electric potential $\phi$ for each phase is obtained similar to Appendix~\ref{app1}. This solution corresponds to a simple case $R=R_1$. In so doing, we get
\begin{eqnarray}
\varphi &=& r(a_i\cos\theta+b_i\sin\theta),\qquad r<R,\\
\nonumber
\varphi \!&=&\!\frac{f_i\cos\theta\!+\!g_i\sin\theta}{r}\!-
  \!\frac{j_0r\left[\sigma_{xx}^e\cos{\theta}\!+\!\sigma_{xy}^e\sin{\theta}\right]}{\sigma_{xx}^{e2}\!+\!\sigma_{xy}^{e2}},\,\,\, r>R.
\end{eqnarray}
The constants $a_i$, $b_i$, $f_i$, and $g_i$ are obtained from the matching $\varphi$ and $j_r$ at $r=R$. The self-consistency condition remains the same, Eq.~\eqref{self_consA}, while for the average electric field now we have
\begin{equation}\label{aver_R}
\!\!\!\!\pi R^2\bar{\bf E}\!=\!p\!\!\int_0^R\!\!\!\!\!rdr\!\!\int_0^{2\pi}\!\!\!d\theta {\bf E}_1(\mathbf{r})+(1-p)\!\!\int_0^R\!\!\!\!\!rdr\!\!\int_0^{2\pi}\!\!\!d\theta {\bf E}_2(\mathbf{r}),
\end{equation}
where ${\bf E_{i}}$ is the electric field in the phase with conductivity tensor $\hat{\sigma}_i$.

After rather cumbersome but straight-forward algebra, we derive equation system for the components of the effective conductivity
\begin{widetext}
\begin{eqnarray}\label{system}
 \frac{p\left(\sigma_{xx}^{(1)}+\sigma_{xx}^e\right)} {\left(\sigma_{xx}^{(1)}+\sigma_{xx}^e\right)^2 + \left(\sigma_{xy}^{(1)}-\sigma_{xy}^e\right)^2} +
 \frac{\left(1-p\right)\left(\sigma_{xx}^{(2)}+\sigma_{xx}^e\right)} {\left(\sigma_{xx}^{(2)}+\sigma_{xx}^e\right)^2 + \left(\sigma_{xy}^{(2)}-\sigma_{xy}^e\right)^2} =
 \frac{1}{2\sigma_{xx}^e},  \\
 \frac{p\left(\sigma_{xy}^{(1)}-\sigma_{xy}^e\right)} {\left(\sigma_{xx}^{(1)}+\sigma_{xx}^e\right)^2+\left(\sigma_{xy}^{(1)}-\sigma_{xy}^e\right)^2} +
 \frac{\left(1-p\right)\left(\sigma_{xy}^{(2)}-\sigma_{xy}^e\right)} {\left(\sigma_{xx}^{(2)}+\sigma_{xx}^e\right)^2 + \left(\sigma_{xy}^{(2)}-\sigma_{xy}^e\right)^2} = 0.
\end{eqnarray}
\end{widetext}

In the limit of low magnetic field, we can neglect the terms with $(\sigma_{xy}^{(i)})^2$, which are quadratic in $B$. In this case we have $R_{\textrm{xy}}=\sigma_{xy}^e/(\sigma_{xx}^e)^2$. The first of Eqs.~\eqref{system} reduces to quadratic one. We solve it and obtain
\begin{widetext}
\begin{eqnarray*}
\sigma_{xx}^e&=&(0.5-p)(\sigma_{xx}^{(2)}-\sigma_{xx}^{(1)})+ \sqrt{(0.5-p)^2(\sigma_{xx}^{(2)}-\sigma_{xx}^{(1)})^2+\sigma_{xx}^{(1)}\sigma_{xx}^{(2)}},\\
\sigma_{xy}^e&=&\frac{p\sigma_{xy}^{(1)}(\sigma_{xx}^{(2)}+\sigma_{xx}^e)^2+(1-p)\sigma_{xy}^{(2)}(\sigma_{xx}^{(1)}+\sigma_{xx}^e)^2}{p(\sigma_{xx}^{(2)}+\sigma_{xx}^e)^2+(1-p)(\sigma_{xx}^{(1)}+\sigma_{xx}^e)^2}.
\end{eqnarray*}
\end{widetext}

\end{document}